\documentclass[useAMS]{mn2e}

\usepackage{ulem} 
\usepackage{graphicx}
\usepackage{amsmath}
\usepackage{ulem}
\usepackage{rotating}
\usepackage{amssymb}

\newcommand{\hst}{{\sl HST}}

\def\lapprox{\hbox{\lower .8ex\hbox{$\,\buildrel < \over\sim\,$}}}
\def\gapprox{\hbox{\lower .8ex\hbox{$\,\buildrel > \over\sim\,$}}}

\title[Extending \textit{Gaia}\,DR2 with \textit{HST} astrometry]{
  Extending \textit{Gaia}\,DR2 with \textit{HST} narrow-field astrometry: the WISE\,J154151.65$-$225024.9 test case\thanks{
Based on observations with the NASA/ESA {\it Hubble
Space Telescope}, obtained at the Space Telescope Science Institute,
which is operated by AURA, Inc., under NASA contract NAS 5-26555.
}
}
\author[L.\,R.\,Bedin]{
  L.\,R.\,Bedin$^{1}$\thanks{E-mail: luigi.bedin@oapd.inaf.it} and 
  C. Fontanive$^{2,3}$
\\  
$^{1}$INAF-Osservatorio Astronomico di Padova, Vicolo dell'Osservatorio 5, I-35122 Padova, Italy\\
$^{2}$Institute for Astronomy, University of Edinburgh, Blackford Hill, Edinburgh EH9 3HJ, UK\\
$^{3}$Centre for Exoplanet Science, University of Edinburgh, Edinburgh EH9 3HJ, UK
}

\begin{document} 

\date{Accepted 2018 September 24. Received 2018 September 24; in original form 
2018 July 11}

\pagerange{\pageref{firstpage}--\pageref{lastpage}} \pubyear{201X}

\maketitle
 
\label{firstpage}

\begin{abstract}
One field containing WISE\,J154151.65$-$225024.9 was observed by
\textit{Hubble Space Telescope} at three different epochs taken in
$\sim$5\,yrs. We measured positions of sources in all images and
successfully linked these positions to the \textit{Gaia}\,DR2 absolute
system to derive the astrometric parameters for this faint close-by Y1
brown dwarf.
The developed procedure avoids traditional limitations of relative
imaging-astrometry with narrow-field cameras, extending Gaia\,DR2 to
fainter magnitudes.
%
%We found $(\alpha,\delta,\mu_{\alpha\cos{\delta}},\mu_\delta,\pi)_{2000.0}=$
%($235^\circ\!\!.4678705\pm5.3$\,mas,
%$-22^\circ\!\!.84020586\pm5.7$\,mas,
We found $(\mu_{\alpha\cos{\delta}},\mu_\delta,\pi)=$
($-902.62\pm0.35$\,mas\,yr$^{-1}$,
$-88.26\pm0.35$\,mas\,yr$^{-1}$,
$168.4\pm2.2$\,mas$)$, which represent a sizable improvement over recent
determinations in the literature.
Applying a correction from relative ($\pi$) to absolute parallax ($\varpi$) we found
$\varpi$=169$\pm$2\,mas, corresponding to a distance of 5.9$\pm$0.1\,pc. 

%
% X(1) =    15.697858030862172     
% X(2) =   -22.840205862620905     
% X(3) =   -902.61562101144432     
% X(4) =   -88.261065374289302     
% X(5) =    168.37633145012094     
%  
%
% ERRORS FROM 25000 REALIZATIONS 
% RAo_2000 =   15.69785803 +/- 5.319858747
% DCo_2000 =  -22.84020586 +/- 5.742136127
% MU_RAcsD = -902.61797440 +/- 0.3549084167
% MU_DECLI =  -88.26248465 +/- 0.3453236302
% PARALLAX =  168.38317780 +/- 2.229892366

\end{abstract}


%%%%%%%%%%%%%%%%%%%%%%%%%%%%%%%%%%%%%%%%%%%%%%%
%
\section{Introduction}
\label{introduction}
%
%%%%%%%%%%%%%%%%%%%%%%%%%%%%%%%%%%%%%%%%%%%%%%%
%https://sites.google.com/view/ydwarfcompendium/y-dwarfs/wise-1541-2250
%
The brown dwarf (BD)
WISE\,J154151.65$-$225024.9\footnote{\texttt{http://simbad.u-strasbg.fr\-/simbad/sim-id?\-Ident=WISE+J154151.65$-$225024.9}}
(hereafter, W1541$-$2250) was discovered by Cushing et al.\ (2011),
who also found it to have a parallactic distance strongly in
disagreement with its spectroscopic and photo-metric values.
The same authors found a spectral type classification Y0 for
W1541$-$2250, which later, Kirkpatrick et al.\ (2012) redetermined as
Y0.5, and finally the new space-based spectra by Schneider et
al.\ (2015) reclassified it as a BD of Y1.
However, Beichman et al.\ (2014) and Schneider et al.\ (2015) found
difficulties in fitting the spectra and the photometry to different
models, obtaining rather unconstrained parameters, especially ages
ranging between 0.6 and 14\,Gyr, masses in the range 12-31\,M$_{\rm
  Jup}$, temperatures of 350-441\,K, but also radii of
0.87-1.0\,R$_{\rm Jup}$, and $\log g$ of 4.-5.
Worth to mention that no binarity was detected for W1541$-$2250 (Opitz
et al.\ 2016), although they ruled out mainly near equal mass
binaries.

The early estimate of the distance (d) for W1541$-$2250 was given in
Kirkpatrick et al.\ (2011), d=$2.8^{+1.3}_{-0.6}$\,pc, in strong
disagreement with their spectro-photometric estimates, which were
placing the BD at 8.2\,pc.
Subsequently Kirkpatrick et al.\ (2012) still found large
discrepancies between the parallax measurement in Kirkpatrick et
al.\ (2011) and other measurements so they derived a
spectro-photometric distance of 4.2\,pc, to which they refer as the
``adopted'' distance.

Others large discrepancies appeared in the literature for the parallax
of this object, noticeably, Dupuy \& Kraus (2013) suggest a parallax
of at most 148\,mas (d$\sim$6.75\,pc), and Marsh et al.\ (2013) giving
d$>$6\,pc.
The discrepancy of parallactic distance with photo-spectroscopic
estimates was recognized by Tinney et al.\ (2014) as partially due to
the angular proximity ($\sim1^{\prime\prime}$) to a much brighter
($\sim$2.5\,mag) field stars just South of W1541$-$2250 at the epoch
of their observations. For this reason Tinney et al.\ (2014) did not
use the declination in the fit of the astrometric parameters.

Parallaxes derived by Tinney et al.\ (2014) and by Beichman et
al.\ (2014) are the most recent and accurate estimates, and they
amount to 175.1$\pm$4.4\,mas and 176$\pm$9\,mas, the latter work also
based on two \textit{Hubble Space Telescope (HST)} epochs.
[Note that while the proximity to a star affected previous works
  carried with \textit{Spitzer} and ground-bases telescopes, it is not
  an issue with the high resolution of \textit{HST}.]
These are currently the best value of parallax available so far,
implying a distance d$\sim$5.71$\pm$0.15\,pc.

In this work we take advantage of the Gaia\,DR2 (Gaia collaboration
2016, 2018) to anchor all existing \hst\ data of W1541$-$2250 to an
absolute reference system, and to independently derive the astrometric
parameters for this object to unprecedented levels of accuracy, thanks
to the homogeneity and to the space-based data-set employed here,
which do not suffer of the usual limitations of ground-based
facilities (Bedin et al.\ 2017).

%%%%%%%%%%%%%%%%%%%%%%%%%%%%%%%%%%%%%%%%%%%%%%%%%%%%
%
\section{Observations} 
%
%%%%%%%%%%%%%%%%%%%%%%%%%%%%%%%%%%%%%%%%%%%%%%%%%%%%

This is an imaging-astrometry investigation and all the images
employed were collected with the \textit{Infra Red} (IR) channel of
the \textit{Wide Field Camera 3} (WFC3) at focus of the \hst.
Two archival epochs both from GO\,12970 (PI: Cushing) were collected
in February 12$^{\rm th}$ and May 9$^{\rm th}$ 2013, and a third
proprietary epoch was taken in February 17$^{\rm th}$ 2018 under
GO\,15201 (PI: Fontanive).

The first epoch consists of one \hst-orbit, split into 4 dithered
images in filter F125W, each image in multiaccum mode (with instrument
parameters \texttt{NSAMP=13, SAMP-SEQ=SPARS50})\footnote{
  WFC3 Instrument Handbook, Sect.\,7.7
  \texttt{http://www.stsci.edu/\-hst/\-wfc3/\-documents/\-handbooks/\-currentIHB/\-c07\_ir08.html}
}  
with a duration of 602.930\,s.
The second epoch has twice as many dithered exposures, but
considerably shallower, made of fewer \texttt{NSAMP} samplings, and
split in two filters: 4 exposures of 77.934\,s in F125W
(\texttt{NSAMP=4, SAMP-SEQ=SPARS25}) and 2 exposure of 102.934\,s and
2 of 77.934\,s in F105W (\texttt{NSAMP=5 and 4, SAMP-SEQ=SPARS25}).
Note that program GO\,12970 also collected many grism spectra, which
we do not use because they are not suitable for imaging-astrometry.
Note that these two \textit{HST} epochs were also analyzed by Beichman
et al.\ (2014).

The third epoch from GO\,15201 consists of 4 well dithered exposures
of 299.232\,s in F127M (\texttt{NSAMP=11, SAMP-SEQ=STEP50}).  Within
this program F139M images were also collected but of not meaningful
signal due to the faintness of BDs in this band (see Fontanive et
al.\ 2018 for the rational behind the F127M/F139M filter choice).
Table\,\ref{tabimg} gives information for all the 16 images used in
this work. 
%

%__________________________________________________________________
%
\begin{table}
  \caption{\textit{HST} images used in this work. Where MJD$_{\rm
      start}$ is the modified Julian day at the start of the exposure,
    EXPT the exposure time duration, and PA$_{\rm V3}$ is the Position
    Angle of axis V3 of the \hst's focal plane (Dressel et al.\ 2017).}
  \center
\begin{tabular}{lcc}
\hline
\#ID: MJD$_{\rm start}$ & image~ EXPT & PA$_{\rm V3}[^\circ]^{(*)}$\\
\hline
 & & \\
 F125W GO\,12970 & epoch 2013.1 & \\
 & & \\
01: 56335.74652883 & \texttt{ic2j09joq}~ 603\,s & 102.8093 \\ 
02: 56335.75422568 & \texttt{ic2j09jpq}~ 603\,s & 102.8094 \\
03: 56335.76192235 & \texttt{ic2j09jrq}~ 603\,s & 102.8094 \\
04: 56335.76961901 & \texttt{ic2j09jtq}~ 603\,s & 102.8093 \\
 & & \\                                                   
 F105W GO\,12970 & epoch 2013.3 & \\                                      
 & & \\                                                   
05: 56421.51200515 & \texttt{ic2j31ssq}~ 103\,s & 147.9995 \\
06: 56421.54665811 & \texttt{ic2j31sxq}~ 103\,s & 147.9995 \\
07: 56421.57539052 & \texttt{ic2j31t5q}~  78\,s & 147.9995 \\
08: 56421.57701107 & \texttt{ic2j31t6q}~  78\,s & 147.9996 \\
 & & \\                                                   
 F125W GO\,12970 & & \\                                      
 & & \\                                                   
09: 56421.68076089 & \texttt{ic2j38tgq}~  78\,s & 147.9995 \\
10: 56421.68238145 & \texttt{ic2j38thq}~  78\,s & 147.9995 \\
11: 56421.68400182 & \texttt{ic2j38tiq}~  78\,s & 147.9995 \\
12: 56421.68562218 & \texttt{ic2j38tjq}~  78\,s & 147.9994 \\
 & & \\                                                   
 F127M GO\,15201 & epoch 2018.1 & \\                                      
 & & \\                                                   
13: 58166.84740268 & \texttt{idl222jdq}~ 299\,s & 104.0010 \\
14: 58166.85160416 & \texttt{idl222jeq}~ 299\,s & 104.0012 \\
15: 58166.85580564 & \texttt{idl222jgq}~ 299\,s & 104.0012 \\
16: 58166.86000712 & \texttt{idl222jiq}~ 299\,s & 104.0010 \\
 & & \\
\hline
\end{tabular}
\label{tabimg}
\end{table} 
\section{Data Reduction and Data Analysis} 
%
%%%%%%%%%%%%%%%%%%%%%%%%%%%%%%%%%%%%%%%%%%%%%%%%%%%%

In the following we will give a brief description on how the raw
positions in pixel coordinates $(x,y)$ for all the sources in the
individual frames were obtained, corrected for distortion, 
%$(x^{\rm cor},y^{\rm cor})$
transformed into a common reference frame
$(X,Y)$, and then transformed into the equatorial coordinate
$(\alpha,\delta)$ of Gaia\,DR2 reference system at
epoch 2013.1. We will give reference to articles containing
detailed descriptions of procedures and software.

%%%%
\subsection{Fluxes and positions in the individual images}
%%%
%
Positions and fluxes of sources in each WFC3/IR \texttt{\_flt} image
were obtained with a software that is adapted from the program
\texttt{img2xym\_WFC.09x10} initially developed for ACS/WFC (Anderson
\& King 2006), and now publicly available for WFC3 too.\footnote{
\texttt{http://www.stsci.edu/$\sim$jayander/WFC3/}
}
Together with the software, a library of \textit{effective}
point-spread functions (PSFs) for most common filters is also
released. These can be perturbed in a spatially variable (up to
3$\times$3) array to better fit PSFs of each individual frame.
These procedures tailor the library PSFs to each individual image even
better than spatially-constant perturbed PSFs, as they better account
for small focus variations across the whole field of view (see
Anderson \& Bedin 2017 for general principles).
In addition to solving for raw positions and fluxes, the software also
provides a quality-of-fit parameter ($Q$). The quality-of-fit
essentially tells how well the flux distribution resembles the shape
of the PSF (this parameter is defined as in Anderson et al.\ 2008).
It is close to zero ($Q<$\,0.03) for stars measured best. This
parameter is useful for eliminating galaxies, blends, and stars
compromised by detector cosmetic or artifacts ($Q>$\,0.3).

Once the raw pixel positions $(x^{\rm raw},y^{\rm raw})$ and magnitude
are obtained, they are corrected for geometric distortion of the
camera. We used the best available average distortion corrections for
WFC3/IR (also derived by Anderson and publicly available$^3$) to
correct the raw positions of sources that we had measured within each
individual image.
We refer to corrected positions of sources in the individual frames
with the symbols $(x^{\rm cor},y^{\rm cor})$.

\subsection{The reference frame $(X,Y)_{2013.1}$}
\label{master}

Among the existing three \hst\, epochs, the one with the highest
signal for W1541$-$2250 is the first ($\sim$2013.1).
All images within this epoch (\#1-4, Table\,\ref{tabimg}) were taken
in a $\sim$45\,minutes time span, and therefore we can safely assume
no (sizable) intrinsic motion of sources observed within this epoch
(we will see the derived motion corresponding to less than 9\,$\mu$as
in 45\,min).

We adopted the $(x^{\rm cor},y^{\rm cor})$ positions of the
best-fitted ($Q<0.2$) sources in image \#4 (\texttt{ic2j09jtq}) as our
initial reference frame ($\sim$250 objects).
Common stars are then used to find the most general linear
transformation (six parameters), between the $(x^{\rm cor},y^{\rm
  cor})_{\#04}$ and the distortion-corrected positions $(x^{\rm
  cor},y^{\rm cor})_{\#0i=1,2,3}$ of the other 3 images in epoch
2013.1.
Next, we use these transformations to compute the clipped mean of the
$(x^{\rm cor},y^{\rm cor})_{\#0i=1,...,4}^{\rm transf.to,\#4}$
positions measured in at least 3 out of the 4 images, and define a
more robust estimate of relative positions for 217 sources.  The
resulting frame of coordinates is indicated with $(X,Y)_{2013.1}$ and
is our adopted reference frame at the reference epoch 2013.1.  Where
\textit{epochs} were computed as Julian years, JYs = 2000.0 + (JD -
2451545.0) / 365.25, where JD = MJD + 2400000.5, and MJD is the
modified Julian day at mid-exposure.

%%%
\subsection{Notation}
%%%
We will indicate the equatorial coordinates in ICRS for a given
\textit{epoch} as $(\alpha,\delta)_{epoch}$.
To transform standard equatorial coordinates $(\alpha,\delta)_{epoch}$
to the pixel coordinates $(X,Y)_{epoch}$ we will make use of the
tangent plane at tangent point $(\alpha_\circ,\delta_\circ)$.
Equatorial coordinates projected on the tangent plane are indicated as
$(\xi,\eta)_{epoch}$.
The most-general linear transformation from $(X,Y)$ to $(\xi,\eta)$ is
a 6-parameters, these transformations will be indicated hereafter with
$\mathcal{A,B,C,D,X_\circ,Y_\circ}$.
To transform $(X,Y)$ into $(\xi,\eta)$ (and visa-versa) we use the
following relations:
\begin{equation}
\begin{cases}
 \xi  = \mathcal{A} (X-\mathcal{X_\circ}) + \mathcal{B} (Y-\mathcal{Y_\circ}) \\
 \eta = \mathcal{C} (X-\mathcal{X_\circ}) + \mathcal{D} (Y-\mathcal{Y_\circ}) \\
\end{cases}
\label{XY2xe}
\end{equation}
\begin{equation}
\begin{cases}
 X = \mathcal{A}^{-1} \xi + \mathcal{B}^{-1} \eta + \mathcal{X_\circ} \\
 Y = \mathcal{C}^{-1} \xi + \mathcal{D}^{-1} \eta + \mathcal{Y_\circ} \\
\end{cases}
\label{xe2XY}
\end{equation}
where the inverse coefficients are derived as:
with $\Delta = \mathcal{(AD-BC)}$.  

To transform $(\xi,\eta)$ to $(\alpha,\delta)$ (and visa-versa) is
more elaborate but remains a classic procedure (e.g., Smart 1931, eq.\,16, 19, 21 and 22).
% This is equation (19) pg. 284 of Smart 1931
\begin{equation}
\begin{cases}
%%%    
  \xi   = \frac{\cos{\delta}\sin{(\alpha-\alpha_\circ)}}
        {\sin{\delta_\circ}\sin{\delta}+\cos{\delta_\circ}\cos{\delta}\cos{(\alpha-\alpha_\circ)}}\\
%%%        
 \eta   = \frac{\cos{\delta_\circ}\sin{\delta}-\sin{\delta_\circ}\cos{\delta}\cos{(\alpha-\alpha_\circ)}}
        {\sin{\delta_\circ}\sin{\delta}+\cos{\delta_\circ}\cos{\delta}\cos{(\alpha-\alpha_\circ)}}\\
%%%
\end{cases}
\label{RD2xe}
\end{equation}
and
\begin{equation}
\begin{cases}
%%%    
  \alpha = \alpha_\circ + {\rm tan}^{-1}(\frac{\xi}
         {\cos{\delta_\circ} - \eta~\sin{\delta_\circ}} )\\
%%%
  \delta = {\rm tan}^{-1}\Bigl\{
%  \cos{(\alpha-\alpha_\circ)}
  \cos{\big[
    {\rm tan}^{-1}( \frac{\xi} {\cos{\delta_\circ} - \eta~\sin{\delta_\circ}} )
    \big]}
  \frac{\sin{\delta_\circ}+\eta\cos{\delta_\circ}}{\cos{\delta_\circ}-\eta\sin{\delta_\circ}}\Bigl\}. 
%
%%%
\end{cases}
\label{xe2RD}
\end{equation}

\subsection{Link $(X,Y)_{2013.1}$ to \textit{Gaia}\,DR2}
\label{link}

There are no such things as reference grids on the sky; in astrometry
we can only measure the source positions registered on detectors,
correct them for instrumental features (such as distortion, etc.), and
compare them with positions observed at other epochs to obtain
transformations that enable us to measure standard coordinates,
motions, and parallaxes.

However, all stars on the sky \textit{do} move --at some level-- and
to correctly transform positions of sources from one frame at a given
epoch to another frame at a different epoch, we first need to know how
stars intrinsically moved on the sky, so as to place them at their
correct positions when observations were made before computing the
transformations.
If the intrinsic motions of stars are not accounted for, these will
affect the accuracy of the transformation of coordinates from an epoch
to another.
Therefore, in addition to the unavoidable measurement errors at any
given epoch, there are also the errors in the transformations (which
are usually derived from the measured positions of a sub-set of
sources at the different epochs).
Transformations and motions of individual stars could be iteratively
solved with difficulties in a stellar field, or ignored if
sufficiently small compared to uncertainties in the positioning (as in
most of the cases).

Thankfully, the \textit{Gaia}\,DR2 not only provides positions at the
reference epoch 2015.5, but also provides individual motions for most
of its sources. Hence, if we choose to refer our positions to sources
in the \textit{Gaia}\,DR2 with motions, we can always know (almost)
exactly where sources were (or will be) at any given epoch,
considerably reducing the errors in the transformations.\\

%%%

In the rest of this work we will derive all our astrometric parameters
of W1541$-$2250 in the observational plane $(X,Y)_{2013.1}$.
Positions of \textit{Gaia}\,DR2 sources are given in equatorial
coordinates on the ICRS for epoch 2015.5, in degrees, so we first
simply correct for proper motions (pms) of the individual sources
using the time base-line, according to:
\begin{equation}
\begin{cases}
\alpha_{epoch} = \alpha_{\rm 2015.5} + \mu_{\alpha^*}/3600/1000 \times (epoch-2015.5) \\
\delta_{epoch} = \delta_{\rm 2015.5} + \mu_\delta/3600/1000 \times (epoch-2015.5) \\
\end{cases}
\label{gaia2ep}
\end{equation}
where $epoch=2013.117719$, and $(\mu_{\alpha^*},\mu_\delta)$ are the
\textit{Gaia}\,DR2 pms in Equatorial coordinates (which are expressed
in mas\,yr$^{-1}$, and where $\alpha^*=\alpha\cos{\delta}$).

Within the studied field of view we found 27 \textit{Gaia}\,DR2 point
sources in common with our reference frame $(X,Y)_{2013.1}$ defined in
Sect.\,\ref{master}.
We adopt as tangent point for the tangent plane
$(\alpha_\circ,\delta_\circ)\equiv(235^\circ\!\!.465250,-22^\circ\!\!.840275)$
and compute the corresponding coordinates on the tangential plane
$(\xi,\eta)_{2013.1}$ using Eq.\,\ref{RD2xe}.
We then solve for $\mathcal{A,B,C,D},X_0,Y_0$ the linear equations at
Eq.\,\ref{XY2xe}, where $(X,Y)_{2013.1}$ are the observables and
$(\xi,\eta)_{2013.1}$ are derived from \textit{Gaia}\,DR2.

These 27 sources had transformed positions consistent to within
$\sim$24\,mas (or within $\sim$0.2 WFC3/IR pixels), 26 within 12\,mas
(0.1\,pixels) and the best 15 of these to better than 3.6\,mas
(0.03\,pixels). We used only these best 15 to calibrate our six linear
terms.
The coefficients of the transformations are given in
Table~\ref{tabRDXY}.
[We will see later how it could be possible to correct from relative
to absolute parallaxes, potentially with an exact correction rather
than statistically.]

\begin{figure}
\begin{center}
\includegraphics[width=88mm]{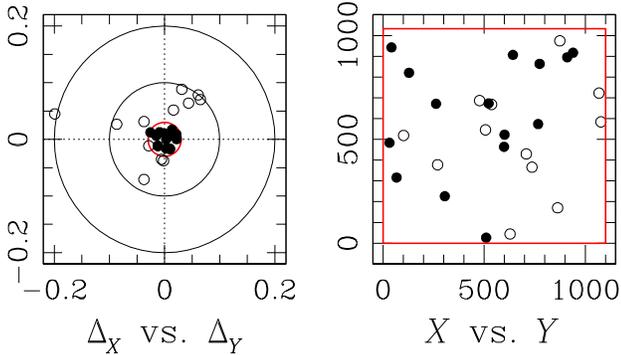}
\caption{
\textit{(Left:) } Consistency in positions ---in units of WFC3/IR
pixels--- of the sources in common between the \textit{Gaia\,DR2}
and our \textit{HST} reference frame. The concentric circles indicate
displacements at 0.03, 0.1 and 0.2\,pixels. The sources with positions
consistent to better than 0.03\,pixels (filled symbols) are used to
define the astrometric reference frame (see text).
\textit{(Right:)} With the same symbols, the spatial distribution of
these sources in the observational plane $(X,Y)$.
\label{go27}
}
\end{center}
\end{figure}

In Fig.\,\ref{go27} we show for these common sources the consistency
in positions (left panel) and their spatial distribution in the
$(X,Y)$ coordinate system (right panel).
In the next sub-section we give a closer look at these 27 sources in
common with \textit{Gaia}\,DR2, investigating the reasons of their
different consistency in positions with \textit{HST} one by one.\\

%__________________________________________________________________
%__________________________________________________________________

%__________________________________________________________________
%
\begin{table}
\caption{Adopted coefficients to transform the observed coordinates
  $(X,Y)_{2013.1}$ into the tangent plane $(\xi,\eta)_{2013.1}$ linked to
  \textit{Gaia}\,DR2 equatorial coordinates
  $(\alpha,\delta)_{2013.1}$.  } \center
\begin{tabular}{lc}
%\hline
\hline
%
% 0.287632961312E-04
% 0.173824063893E-04
% 0.173913255399E-04
%-0.287555630196E-04
%0.570630651526E+03
%0.564770033905E+03
% SCALE 0.120983990267E+00
%# RA_CEN:   235.465250
%# DE_CEN:   -22.840275
%#    RAD:     0.030000
%#  EPOCH: 2013.117719d0
$\mathcal{A}$      & $   ( 0.2876330\pm 0.0025)E-04 $  \\
$\mathcal{B}$      & $   ( 0.1738241\pm 0.0025)E-04 $  \\
$\mathcal{C}$      & $   ( 0.1739133\pm 0.0025)E-04 $  \\
$\mathcal{D}$      & $   (-0.2875556\pm 0.0025)E-04 $  \\
$X_0$      & $570.631 \pm 0.004$ \\
$Y_0$      & $564.770 \pm 0.004$ \\
$\alpha_0$ & $235.465250$ (defined)       \\ 
$\delta_0$ & $-22.840275$ (defined)       \\
\hline
\end{tabular}
\label{tabRDXY}
\end{table} 
%__________________________________________________________________
%
%

The plate-scale derived from \textit{Gaia}\,DR2 of our reference frame
$(X,Y)_{2013.1}$ is
$\sqrt{|{\mathcal{A}\mathcal{D}-\mathcal{B}\mathcal{C}}|}=\sqrt{|\Delta|}=120.984$\,mas,
in agreement with previous WFC3/IR determinations.\\

%__________________________________________________________________
\begin{figure*}
\begin{center}
\includegraphics[width=135mm]{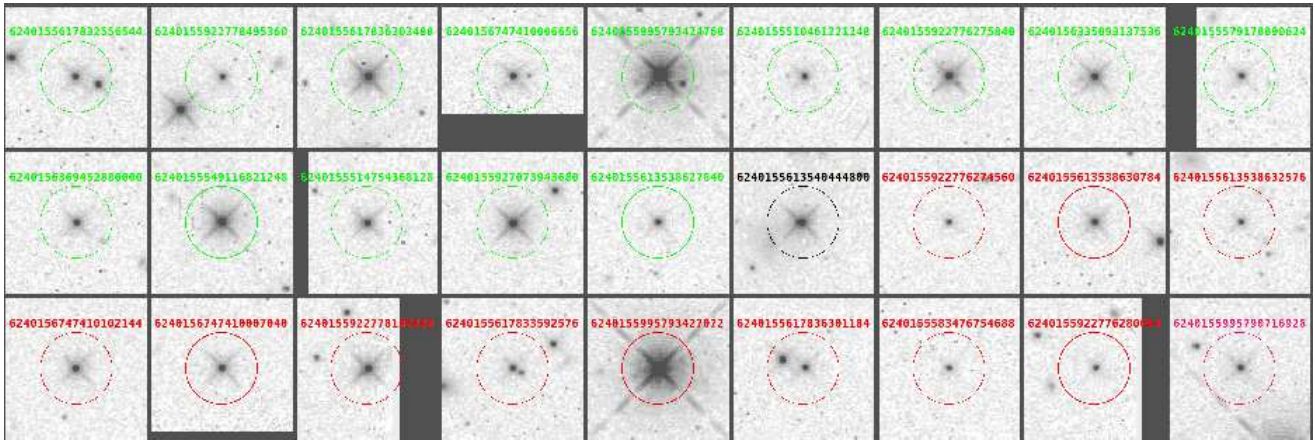}
\caption{
  We extracted from the stack in filter F127M, stamps of
  100$\times$100\,WFC3/IR pixels
  ($\sim$12$^{\prime\prime}$$\times$12$^{\prime\prime}$) centered on
  the 27 sources in common between the \textit{Gaia}\,DR2 catalog and
  our \textit{HST} reference frame.
  Circles with radius of $3^{\prime\prime}$ and labels with the
  \texttt{Identifier:\,Gaia\,DR2} are given.
  From left-to-right and top-to-bottom sources are sorted for
  consistency in positions $\varrho$ (see Table\,\ref{GaiaSs}). The
  best stars, the astrometric references, used to compute the
  transformation in Table\,\ref{tabRDXY}, are labeled in green. In
  black, source \texttt{Gaia\,DR2\,\#6240155613540444800} with a
  significant parallax, and in red stars with displacements between
  0.03 and 0.1\,pixels (3.6-12\,mas).  In magenta the worst source, at
  more than 0.2\,pixels (or 24\,mas). [See Sect.\,\ref{stacks}.]
\label{figX}
}
\end{center}
\end{figure*}
%__________________________________________________________________
%
%

%%%%%%%%%%%%%%%%%%%%%%%%%%%%%%%%%%%%%%%%%%%%%%%%%%%%
%
\subsection{A closer look at the \textit{Gaia}\,DR2 sources}
%
%%%%%%%%%%%%%%%%%%%%%%%%%%%%%%%%%%%%%%%%%%%%%%%%%%%%

More than a third of the sources in common between \textit{HST} and
\textit{Gaia\,DR2} (12/27) have much larger residuals than the others.
Under suggestion of our referee, here we investigate and discuss in
details possible reasons for this.\\

The very first step is to have a visual inspection of the astronomical
scene and search for possible sources compromised by: crowding,
blends, multiple hits of cosmic-rays/bad-pixels, diffraction spikes,
and artifacts in general.
For this purpose, we use the stack (which will be later described in
Sect.\,\ref{stacks}) obtained for images in filter F127M
(Fig.\,\ref{figX}).
We notice that some of the sources with poor positional-consistency
(circled in red) have relatively bright sources closer than
3$^{\prime\prime}$, but we ruled this out as a possible reason, as
sources with the best agreement have similar or even worse neighbors.
Similarly, being close to the boundary of images does not seem to be a
problem, as three sources appear closer than 6$^{\prime\prime}$ to
image boundaries, both in cases of stars labeled in green and in red.
Also, sources in red and green seem to have similar ranges of
brightness.

The spatial distribution of sources in the right-panel of
Fig.\,\ref{go27} seems to suggest that no significant bias of the best
vs.\ worst sources is evident in the plane $(X,Y)$; particularly in
the region around the target, at $\sim$(500,500).
However, given the small number of sources in the field
sufficiently-bright to be detected by \textit{Gaia}, this is not the
ideal case for such a study.

Obviously, the large number of sources with relatively large residuals
(open symbols in the right-panel of Fig.\,\ref{go27}) are not
necessarily due to problems in the \textit{Gaia\,DR2} sources, but
could be in our own measurements in the \textit{HST} images; or in
both.\\

We therefore look at the consistency parameters obtained for sources
in the \textit{HST} reference frame $(X,Y)_{2013.1}$ defined in
Sect.\,\ref{master}.
We look at the trends in the r.m.s.\ of $X$, $Y$, $M_{\rm
  F125W}^{instr.}$ (the instrumental magnitude in F125W), and $Q$ (the
PSF-quality-fit, as defined in Anderson et al.\ 2008), as function of
magnitude.
We found that the brightest star in the field,
\texttt{Gaia\,DR2\,\#6240155995793427072}, has a much larger r.m.s. in
the $X$-position measured within the first epoch (0.148\,pixels)
compared to the r.m.s. of similar magnitude objects (for example, the
second brightest source in the field has an r.m.s.\ of 0.005\,pixels).
This explains the rejection of
\texttt{Gaia\,DR2\,\#6240155995793427072} from the sample of
astrometric reference.
We also noted a significantly larger photometric variability of the
rejected source \texttt{Gaia\,DR2\,\#6240155613538632576}
($\sim$0.055\,mag) compared to that of sources at similar magnitude
($\sim$0.01\,mag), which might potentially indicate it to be an
unresolved binary.\\

%%%

The next step is to consider astrometric and photometric parameters in
the \textit{Gaia\,DR2} catalog.
We inspected for all objects the quality indicators available in
\textit{Gaia}\,DR2\footnote{\texttt{https://gea.esac.esa.int/archive/documentation/GDR2/}}.
We immediately spot two objects with no proper motions \textit{at all}.
These are \texttt{Gaia\,DR2\,\#6240155617836301184} and the
one with the worst positional consistency, i.e.,
\texttt{Gaia\,DR2\,\#6240155995790716928} (in magenta in
Fig.\,\ref{figX}). Evidently, the positions for these objects were not
corrected in Eq.\,\ref{gaia2ep}, and their large displacements might
just reflect their unknown pms.
We also note that \texttt{Gaia\,DR2\,\#6240155613540444800} was
rejected as the only source with a significant parallax greater than
3\,mas.
Of the remaining 9 objects with poor consistency in position in
Fig.\,\ref{go27} (larger than 0.03\,pixels, or 3.6\,mas) we note that
only 4 sources passed the tests for well-measured objects in the
\textit{Gaia\,DR2} catalog as defined in Eqs.\,(C.1) and (C.2) by
Lindegren et al.\,(2018).
These are  \texttt{Gaia\,DR2\,
  \#6240155613538630784, 
  \#6240156747410007040, 
  \#6240155583476754688}, and  
  \texttt{\#6240155922776280064}, with $G$-magnitudes of about 18.8, 19.5, 20.5, and 20.1, respectively.
The expected errors for objects in this magnitude interval are about
2\,mas (Lindegren et al.\ 2018), therefore, the observed positional
residuals of \texttt{Gaia\,DR2} vs.\ \textit{HST} (between 3.6 and
12\,mas) indicate that these four sources could be reasonably well
measured ---given their faintness.  They were rejected simply because
the sum of their random errors (\textit{HST}$\sim$4\,mas and
\textit{Gaia\,DR2}$\sim$2\,mas) are not consistent ($\sim$4.5\,mas)
with our arbitrary tight cut at 0.03\,pixels (3.6\,mas).

In Table~\ref{GaiaSs} we list all the relevant \textit{HST} and
\textit{Gaia\,DR2} parameters for the 27 objects in common. The first
14 are the ones defining the astrometric references, the $15^{\rm th}$
is the one rejected for sizable parallax, the next 11 are the ones
with poor positional consistency, and at last, the one with the worst
consistency ($>$0.2\,pixels).
In the first column we give the \texttt{Identifier:\,Gaia\,DR2}.  Next
column give the parameters from \textit{HST} data: $(X,Y)$-positions,
instrumental magnitude, and $Q$.  Then we give the consistency in
positions $\varrho$ in units of WFC3/IR milli-pixels.  The next
columns give the \textit{Gaia} magnitudes $G$, proper motions in
mas\,yr$^{-1}$, and parallaxes in mas.  The last column gives the most
relevant \textit{Gaia\,DR2} parameters for the present work.
These are:
the \texttt{visibility\_periods\_used} ($v$),
the \texttt{astrometric\_n\_bad\_obs\_al} ($b$), and
the quantities \texttt{U} and \texttt{G} defined
in Eqs.\,(C.1) and (C.2) by Lindegren et al.\ (2018). 
  
%%%----------------------------------------------------------------------------------------
\begin{table*}
	\centering
	\caption{
          The 27 \textit{Gaia\,DR2} sources in common with our
          \textit{HST} master frame. Columns give:
          \texttt{Identifier:\,Gaia\,DR2}, $(X,Y)$-positions in
          WFC3/IR-pixels, \textit{HST}-instrumental magnitude, the
          consistency in positions $\varrho$ in WFC3/IR milli-pixels, the
          \textit{Gaia} magnitude $G$, proper motions in
          mas\,yr$^{-1}$, parallaxes in mas, and four of the most
          significant \textit{Gaia\,DR2} parameters for the present
          work. The visibility periods used $v$, the number of bad
          astrometric observations $b$, and the quantities \texttt{U}
          and \texttt{G} defined in Eqs.\,(C.1) and (C.2) by Lindegren
          et al.\ (2018). The first 14 objects are those used to define the
          astrometric reference, and have all $\varrho<23.3$\,milli-pixel ($\sim$2.8\,mas). }
	\label{GaiaSs}
	\begin{tabular}{cccccc} 
\hline
\texttt{Identifier:\,Gaia\,DR2}
                             & $(X,Y,M_{\rm F125W}^{\textit{instr.}})$
                                                            & $\varrho$ & $G$ 
& ($\mu_{\alpha^*}\pm\sigma_{\mu_{\alpha^*}}$,$\mu_{\delta}\pm\sigma_{\mu_{\delta}}$,$\pi\pm\sigma_\pi$) 
                    & \texttt{($v$,$b$,U,G)}\\
%[\#] & [WFC3/IR-pixels,mag] & [mpppxl]& [mag]& [mas/]& []& []& \\
%(1) & (2) & (3) & (4) & (5) & (6) & (7) & (8)\\                                                                                                                                                         
\hline
%#.......................................................#..... ........ ..... ........ ..... ........ ..... ..... .... ... . 
%# Identifier:GaiaID      X        Y	  MAG    dR     #   GMAG     muRA emuRA    muDEC emuDE      PI    ePI   NVI NBAD   U G
%#        (#)          (IRpxl)   (IRpxl)  (mag)  (IRpxl) #  (mag) (mas/yr) (m/y) (mas/yr) (m/y)    (mas) (mas)                   
%#.......................................................#...... ........ ..... ........ ..... ........ ..... ..... .... . . . 
\texttt{6240155617832556544} & (0522.088,673.142,$-$7.4134) &  2.9 &  19.525 & ($-$8.202$\pm$1.126,$-$6.820$\pm$0.802,$-$0.045$\pm$0.620)&(10,2,1,1)\\
\texttt{6240155922778495360} & (0939.610,917.019,$-$6.8337) & 10.5 &  19.812 & ($-$3.942$\pm$1.127,$-$2.834$\pm$0.753,$-$0.042$\pm$0.648)&(10,1,1,1)\\
\texttt{6240155617836303488} & (0600.795,522.307,$-$9.3597) & 11.7 &  17.983 &($-$10.407$\pm$0.323,$-$1.714$\pm$0.221,$+$0.639$\pm$0.188)&(10,0,1,1)\\
\texttt{6240156747410006656} & (0509.842,026.488,$-$7.6669) & 15.3 &  19.307 & ($-$8.658$\pm$0.949,$-$2.836$\pm$0.821,$+$1.048$\pm$0.500)&(09,1,1,1)\\
\texttt{6240155995793424768} & (0766.966,573.666,$-$12.0318)& 15.6 &  14.418 &($+$11.261$\pm$0.066,$-$12.398$\pm$0.047,$+$0.975$\pm$0.037)&(10,1,1,1)\\
\texttt{6240155510461221248} & (0128.497,820.398,$-$6.8964) & 16.4 &  20.738 & ($+$4.171$\pm$3.918,$-$11.174$\pm$2.764,$-$0.233$\pm$1.700)&(09,0,0,0)\\
\texttt{6240155922776275840} & (0774.182,863.455,$-$8.8226) & 16.4 &  17.811 & ($+$1.576$\pm$0.295,$-$4.212$\pm$0.205,$+$0.058$\pm$0.170)&(10,0,1,1)\\
\texttt{6240156335093137536} & (0066.905,316.118,$-$8.3823) & 16.6 &  18.126 &($-$11.804$\pm$0.365,$-$7.996$\pm$0.249,$-$0.007$\pm$0.206)&(10,1,1,1)\\
\texttt{6240155579178890624} & (0031.905,483.388,$-$6.8282) & 17.2 &  19.720 & ($+$2.040$\pm$1.140,$-$8.564$\pm$0.774,$+$0.158$\pm$0.650)&(10,0,1,0)\\
\texttt{6240156369452880000} & (0305.537,226.296,$-$7.6479) & 18.8 &  18.914 & ($-$3.309$\pm$0.612,$+$0.946$\pm$0.411,$-$0.124$\pm$0.361)&(10,0,1,1)\\
\texttt{6240155549116821248} & (0642.303,906.672,$-$9.6141) & 20.1 &  17.196 & ($-$4.379$\pm$0.221,$-$7.765$\pm$0.147,$+$0.655$\pm$0.121)&(10,0,1,1)\\
\texttt{6240155514754368128} & (0041.535,942.644,$-$7.9155) & 21.9 &  20.537 &($-$21.354$\pm$2.412,$-$0.103$\pm$1.657,$+$1.602$\pm$1.139)&(09,0,1,0)\\
\texttt{6240155927073943680} & (0910.703,895.041,$-$9.2091) & 21.9 &  17.326 &($-$10.000$\pm$0.243,$+$3.948$\pm$0.170,$+$0.217$\pm$0.135)&(10,0,1,1)\\
\texttt{6240155613538627840} & (0262.163,671.320,$-$7.0115) & 23.2 &  20.117 & ($-$1.775$\pm$1.412,$-$10.251$\pm$0.950,$-$0.762$\pm$0.801)&(10,0,1,1)\\
%######                                                                                                                                                           
& & & & &  \\                                                                                                                                                    
\texttt{6240155613540444800} & (0598.187,464.043,$-$9.0561) & 28.5 &  19.354 &($-$37.833$\pm$0.889,$-$25.853$\pm$0.610,$+$3.857$\pm$0.606)&(10,0,1,1)\\
& & & & &  \\                                                                                                                                                    
%######                                                                                                                                                                 
\texttt{6240155922776274560} & (0874.306,974.415,$-$6.1297) & 30.8 &  20.614 & ($+$2.012$\pm$2.491,$-$6.391$\pm$1.710,$+$0.427$\pm$1.166)&(09,0,1,0)\\
\texttt{6240155613538630784} & (0477.394,686.589,$-$7.7311) & 35.6 &  18.839 & ($-$3.394$\pm$0.705,$-$4.478$\pm$0.623,$+$0.512$\pm$0.381)&(09,0,1,1)\\
\texttt{6240155613538632576} & (0269.716,377.640,$-$6.3762) & 37.7 &  20.137 & ($-$6.979$\pm$1.499,$-$5.074$\pm$1.025,$-$0.398$\pm$0.869)&(10,1,1,0)\\
\texttt{6240156747410102144} & (0863.329,170.568,$-$7.9744) & 48.8 &  20.194 &($-$16.071$\pm$2.526,$-$1.265$\pm$1.739,$+$2.181$\pm$0.961)&(07,3,1,1)\\
\texttt{6240156747410007040} & (0628.193,044.235,$-$7.8694) & 54.2 &  19.479 & ($-$0.796$\pm$0.899,$-$11.031$\pm$0.600,$+$0.820$\pm$0.498)&(10,0,1,1)\\
\texttt{6240155922778106880} & (1077.511,584.057,$-$8.4938) & 77.3 &  19.536 & ($-$5.770$\pm$1.063,$-$6.684$\pm$0.727,$+$1.444$\pm$0.621)&(10,3,0,0)\\
\texttt{6240155617833592576} & (0506.614,545.481,$-$7.0274) & 80.0 &  20.525 & ($-$0.175$\pm$2.270,$-$4.630$\pm$1.558,$+$1.224$\pm$1.104)&(09,0,1,0)\\
\texttt{6240155995793427072} & (0736.691,366.113,$-$12.5981)& 90.7 &  14.411 &($-$13.740$\pm$0.072,$+$4.197$\pm$0.048,$+$0.184$\pm$0.041)&(10,0,1,1)\\
\texttt{6240155617836301184} & (0536.728,667.953,$-$7.0182) & 93.8 &  21.160 & ($+$0.000$\pm$0.000,$+$0.000$\pm$0.000,$+$0.000$\pm$0.000)&(06,0,1,0)\\
\texttt{6240155583476754688} & (0101.508,518.783,$-$6.0182) & 95.7 &  20.546 & ($-$9.212$\pm$2.330,$-$1.858$\pm$1.687,$+$1.427$\pm$1.558)&(09,0,1,1)\\
\texttt{6240155922776280064} & (1067.981,722.722,$-$6.3521) & 99.4 &  20.089 & ($-$4.689$\pm$1.359,$-$0.037$\pm$0.935,$-$1.092$\pm$0.797)&(10,0,1,1)\\
%######                                                                                                                                                                         
& & & & &  \\                                                                                                                                                            
\texttt{6240155995790716928} & (0708.427,429.764,$-$7.1248) &204.1 &  20.927 &  ($+$0.000$\pm$0.000,$+$0.000$\pm$0.000,$+$0.000$\pm$0.000) &(06,0,1,0)\\
%%%%%%%%%%%%%%
		\hline
	\end{tabular}
\end{table*}
%%%----------------------------------------------------------------------------------------

%%%%%%%%%%%%%%%%%%%%%%%%%%%%%%%%%%%%%%%%%%%%%%%%%%%%
%
\subsection{Geometric-distortion solution of WFC3/IR}
%
%%%%%%%%%%%%%%%%%%%%%%%%%%%%%%%%%%%%%%%%%%%%%%%%%%%%

In this section we investigate $(i)$ at what level the imperfections
in the adopted WFC3/IR distortion solution could affect our results,
and $(ii)$ whether \textit{Gaia\,DR2} could actually be used to
improve the current geometric distortion solutions of \textit{HST}
cameras.

\subsubsection{Effects of WFC3/IR distortion uncertainties}

First of all, even from frame to frame ---consecutively taken--- the
simple \textit{velocity aberration} (Cox \& Gilliland 2003) can cause
sizable changes in the plate-scale. Good telemetry (or accurate
modelling of the \textit{HST} orbits) can fix these effects (e.g.,
Fig.\,9 in Bellini, Anderson \& Bedin 2011). Note that this is merely
a scale factor induced by the motion of the telescope in special
relativity and has nothing to do with the distortion of the camera
itself, but still is a sizable effect that needs to be taken carefully
into account when applying the geometric distortion solution and
dealing with absolute astrometry.

Second, as extensively discussed in Bedin et al.\ (2014), any
adopted geometric distortion correction for any of the \textit{HST\,}
cameras is just an average solution. 
Even after correction for velocity aberration, there are always
sizable changes and these are mainly induced by focus variations, the
so called \textit{breathing} of the telescope tube, which are the
result of the different incidence of the light from the Sun.
Detailed models of these changes in the geometric distortion with the
focal length are still not developed, so far there are only early
attempts of modelling the changes in the PSFs as function of focal
length (Anderson \& Bedin 2017).

Third,  it  is known  that  linear  terms  of the  ACS/WFC  distortion
solution have  been changing slowly  over time (Anderson  \& Rothstein
2007; Ubeda,  Kozhurina-Platais \& Bedin  2013), the reasons  of these
effects are still  not clear, probably the result of  a slow out gas of
metal that shrink structures and cause these long-term changes.
However,  in  the  case  of  WFC3/IR,  a  recent  study  by  McKay  \&
Kozhurina-Platais et  al.\ (2018) showed  that the linear terms  of the
geometric distortion is  stable at the level of 13\,mas  over an eight
years time span.

The best available geometric distortion solutions for \textit{HST}
cameras are composed by the sum of a polynomial and of an empirical
look-up table (e.g., Anderson \& King 2000, 2003, 2006 for
descriptions in great details).
Thankfully, all of the three effects, i.e., velocity aberration,
breathing, and shrinking, just described, cause detectable changes
\textit{only} in the linear part of the geometric distortion, while
the non-linear part of the distortion seems to remain unchanged within
the uncertainties (at the $\sim$\,1\,mas level).\\

The beauty of our approach is that the six-parameters linear
transformation derived in Sect.\,\ref{master} that calibrates our
\textit{HST} master frame to the absolute reference frame of
\textit{Gaia\,DR2} (Table\,\ref{tabRDXY}), naturally absorbs all of
these three effects, fixing the values of: absolute scale, rotation,
shifts and the two skews terms.
In other words, the linear terms of the geometric distortion can
change by a lot (even by few dozen of mas), but as long as we use the
most-general linear transformation (a 6-parameters) to calibrate our
$(X,Y)$ reference system to \textit{Gaia\,DR2} absolute astrometric
reference frame, all of these effects do not prevent us from achieving
accuracies limited just by our positional accuracy ($\sim$1\,mas) and
by the stability of the non-linear part of the adopted geometric
distortion solution.

Our adopted WFC3/IR geometric distortion solution was derived by
Anderson (2016, pg.\ 39, Appendix\,A).\footnote{Publicly available at
%Instrument Science Report WFC3 2016-12
%http://www.stsci.edu/hst/wfc3/documents/ISRs/WFC3-2016-12.pdf
%http://www.stsci.edu/hst/wfc3/
  \texttt{http://www.stsci.edu/\-$\sim$jayander/\-WFC3/\-WFC3IR\_GC/}
%http://www.stsci.edu/$\sim$jayander/WFC3/WFC3IR\_PSFs/
}
Unlike WFC3/UVIS and the ACS channels, which have separate solutions
for each filter, there is only one solution that works for all WFC3/IR
filters. So far variation with filters were not explored, but any
variation is likely less than $\sim$0.01 pixel (i.e., $\sim$1.2\,mas,
cfr.\ Anderson 2016).
Precisions of $\sim$1\,mas (differential astrometry) with WFC3/IR are
now routinely reached (e.g., Anderson 2016, Bellini et al.\ 2017,
2018, Libralato et al.\ 2018), and the \textit{Gaia\,DR2} catalog now
makes it possible to transform those precisions into
absolute-astrometric accuracies, for the reasons just outlined.
This is demonstrated by the displacements observed in Fig.\,\ref{go27}
(and in a similar figure presented later in Sect.\,\ref{lnk}) of this
work.
%__________________________________________________________________
\begin{figure*}
\begin{center}
\includegraphics[width=150mm]{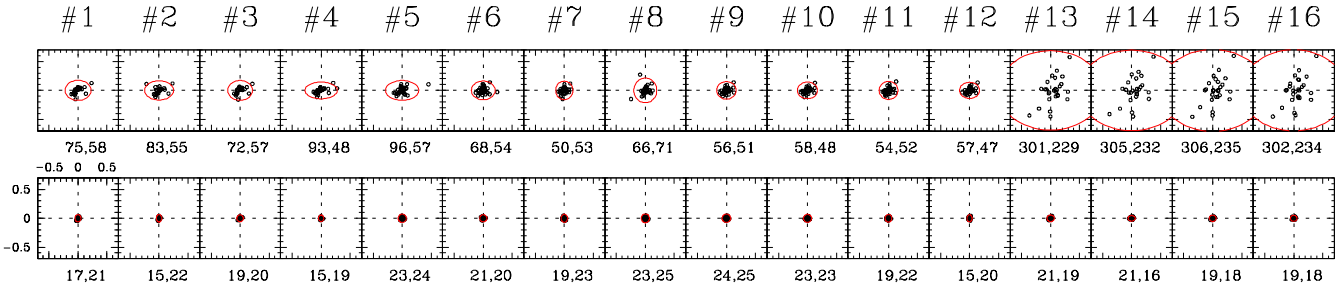}
\caption{
  The sources in common between the \textit{Gaia}\,DR2 catalog and
  those measured (and corrected for distortion) on a given image
  define a six-parameters linear transformation. With this
  transformation we can predict the positions of \textit{Gaia}\,DR2
  sources in that image, and visa-versa.
  Panels in this figure all have the same scale ($-$0.5,+0.5 WFC3/IR
  pixels in both $X$ and $Y$) and show the displacements between the
  observed and the predicted \textit{Gaia}\,DR2 positions based on the
  transformations obtained for each of the 16 images.
  On the \textit{--Top Panels--} however, positions of
  \textit{Gaia}\,DR2 sources were \textit{not} corrected for their
  individual tabulated \textit{Gaia}\,DR2 proper-motions.
  Above each panel the \#id to which each image refers is indicated
  (following notation in first column of Table\,\ref{tabimg}). Note
  that \#1-4 were taken at 2013.1, \#5-12 at 2013.3, and \#13-16 at
  2018.1.
  Red ellipses indicate the 3$\sigma$ of the distribution along the X
  and Y axes, and their 1$\sigma_{X,Y}$ values in WFC3/IR milli-pixels
  are indicated below of each panel.
  On the \textit{--Bottom Panels--} positions of \textit{Gaia}\,DR2
  sources \textit{were} corrected for their individual proper-motions.
  It is clear that in the bottom panels the dispersion of observed
  vs. \textit{Gaia}\,DR2 positions are much tighter (better than
  $\sim$3\,mas) than in the top panels, and also consistent across all
  16 images (with a slight broadening for images with the shortest
  exposure times, particularly in the F105W filter).
\label{figmat}
}
\end{center}
\end{figure*}
%__________________________________________________________________
%
%

In summary, as the non-linear part of the adopted geometric distortion
solution is as accurate as our positional accuracy for best measured
stars, i.e., $\sim$1\,mas, we do not expect residuals in the
distortion to be a significant source of uncertainty in deriving the
astrometric parameters for W1541$-$2250. This target is a source more
than 6\,magnitudes fainter than best measured stars in the field,
having an estimated positional precision between 5 and 12\,mas in
individual images (depending on filter/exposure-time).  Furthermore,
any of such distortion residual would be suppressed by averaging over
multiple dithered observations within each of the individual epochs.

%%%%%%%%%%%%%%%%%%%%%%%%%%

\subsubsection{Can \textit{HST} distortion be improved with \textit{Gaia\,DR2}?}

For a well-exposed point-source the positional accuracy on images from
\textit{HST}' cameras is $\sim$1\,mas (at best 0.32\,mas for
WFC3/UVIS, Bellini et al.\,2011), and the non-linear part of the
geometric distortion solution is proven to be at least that good.

So, even with an infinitely accurate geometric distortion solution for
\textit{HST} cameras, a given target would require multiple
observations to improve the astrometry at the sub-mas level, given the
limit on the accuracy set by random noise (i.e., the possible exposure
time or saturation level) in individual images.
This drastically reduces the number of applications where a sub-mas
level accuracy for the geometric distortion solution would be worth
and useful.
For example, the comparison with \textit{Gaia\,DR2} in the case of a
group of point sources within an \textit{HST} field of view (such as
stars in a star cluster) can statistically highlight sub-mas trends in
the geometric distortion solution.  In this regard, the recent work by
Kozhurina-Platais et al.\ (2018) discuss the potential importance of
such improvements, especially in the low-order components of the
distortions down to the level of 0.5\,mas or better.\\

However, for high-order components (i.e., on small spatial scales)
there is a fundamental limitation in using \textit{Gaia} astrometric
catalogs to improve \textit{HST}' camera distortion solutions, and
this is the spatial density of \textit{Gaia} sources.
The \textit{Gaia\,DR2} catalog has an 'all-sky'-average density of
$\sim$10 sources per arcmin square, and this essentially is set by its
magnitude limits.
For example, effects such as the WFPC2 34$^{\rm th}$-raw feature
discovered and fixed by Anderson \& King (1999), or the WFC3/UVIS
lithographic signatures found by Kozhurina-Platais et al.\ (2010) and
later characterized in Bellini et at.\ (2011) would have been very
hard to calibrate by using just \textit{Gaia\,DR2} sources.

One might think of comparing \textit{Gaia\,DR2} sources with multiple
fields and multiple epochs or even with the entire archive of
\textit{HST} observations to characterize such effects or any
high-order components of the geometric distortion in general; however
not using the same sources with a suitable density and within the same
epochs, would exponentially complicate the calibration, which also,
might well be variable in time at the sub-mas level.
Self calibration of \textit{HST} (e.g., Anderson \& King 1999, 2000,
2003, 2006, Bellini et al.\ 2011) and local transformations in dense
fields (e.g., Bedin et al.\ 2003, 2014, Anderson \& van der Marel
2009, Bellini et al. 2017, 2018) would offer a much easier way to
calibrate and characterize such effects and high-order components of
the geometric distortion in general, than using sources in common with
\textit{Gaia}.\\

Finally, we want to note that at the sub-mas level ---even ignoring
chromatic effects--- there is always an interplay between the adopted
geometric distortion and the exact shape of the PSFs, as a slightly
different PSFs cause slightly different centroids offsets (which
depend on the adopted PSF-centroid normalization), which result in
slight changes in the distortion.  In other words, at sub-mas level
the adopted PSFs model and the adopted geometric distortion solution
become more and more an indissoluble pair.

%__________________________________________________________________
\begin{figure*}
\begin{center}
\includegraphics[width=88mm]{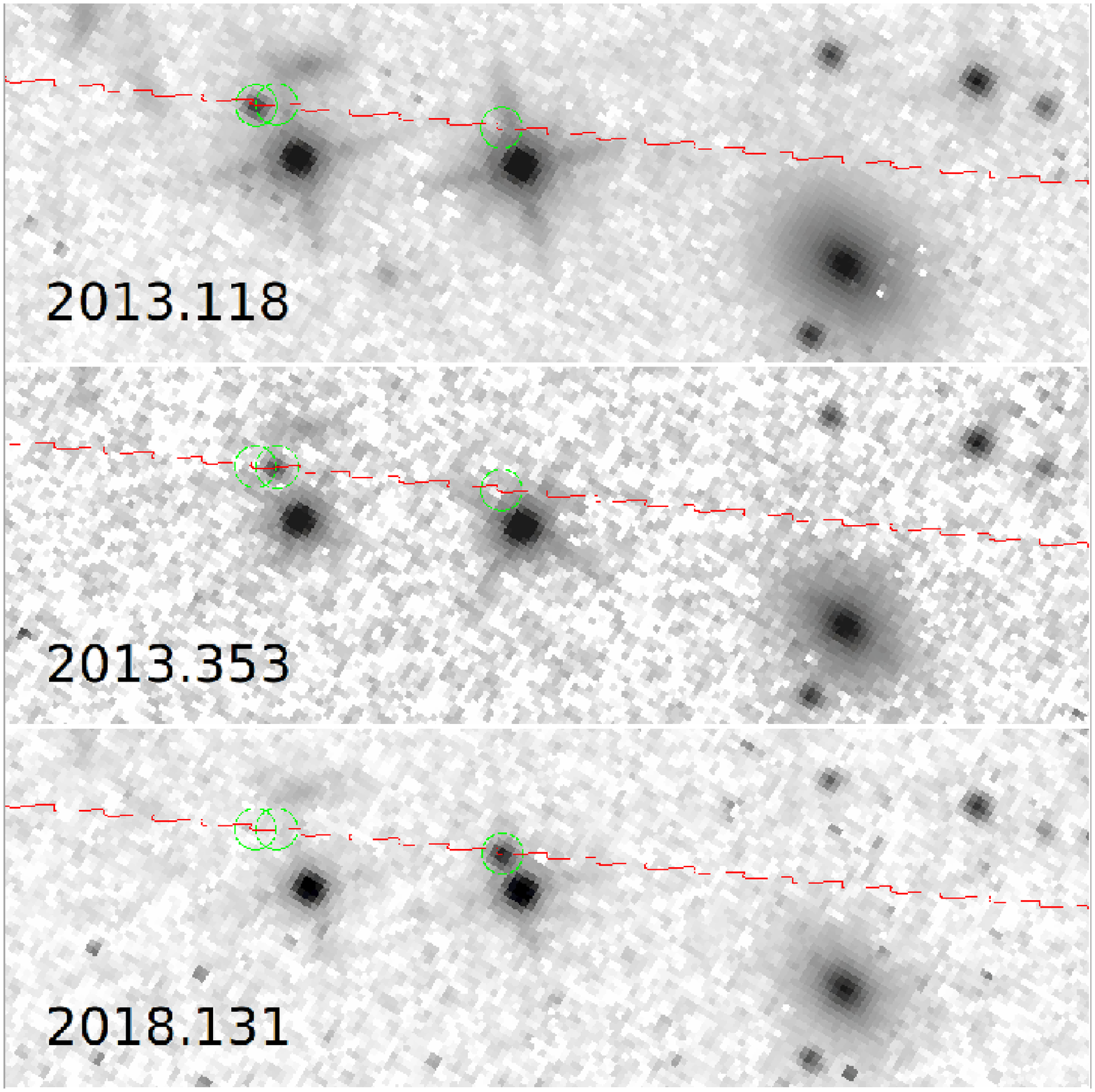}
\includegraphics[width=88mm]{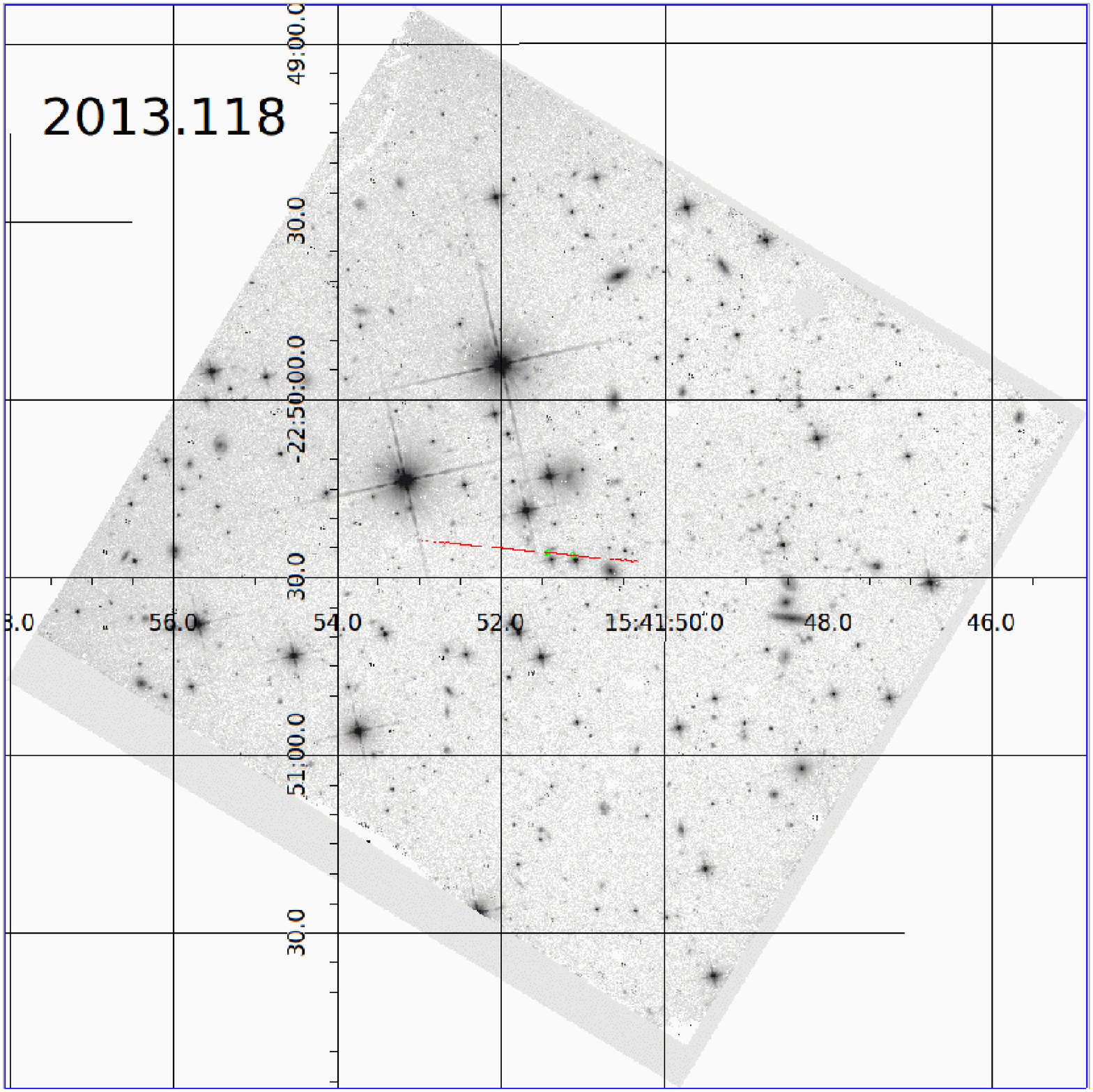} 
\caption{
  \textit{(Left:)} Zoom-in of the \hst\ field surrounding
  W1541$-$2250, as collected in the three epochs analyzed in this
  work.  This small region has a size of $\sim
  20^{\prime\prime}\times6.5^{\prime\prime}$.  Green circles indicate
  the BD positions at the three epochs.  A red line show our
  astrometric solution of the motions for this object in years from
  1990 to 2030.
   \textit{(Right:)} The entire field of view is about
   2.2$^{\prime}$$\times$2.2$^{\prime}$ and this is the stack of four
   WFC3/IR/F125W images collected in first epoch, where the BD has
   highest signal.  The grid and labels are in equatorial coordinates.
\label{stack}
}
\end{center}
\end{figure*}
%__________________________________________________________________

%%%%%%%%%%%%%%%%%%%%%%%%%%%%%%%%%%%%%%%%%%%%%%%%%%%%
%
\subsection{Link all epochs to \textit{Gaia}\,DR2}
%
%%%%%%%%%%%%%%%%%%%%%%%%%%%%%%%%%%%%%%%%%%%%%%%%%%%%
\label{lnk}

Similarly to what was done for the reference epoch, we can use
Eq.\,\ref{gaia2ep} to have pms-corrected position of
\textit{Gaia}\,DR2 sources at all the different epochs of images in
Table\,\ref{tabimg}.
Again, Eq.\,\ref{RD2xe} is used to have the coordinates in the
tangential plane $(\xi,\eta)_{epoch}$.
However, this time we will not have to re-derive the coefficients of
the linear transformation from \textit{Gaia} $(\alpha,\delta)_{epoch}$
to $(X,Y)_{2013.1}$, as those were determined in Sect.\,\ref{link}.
We now have the \textit{Gaia}\,DR2 positions of individual sources,
corrected for their peculiar motions, at any given epoch, and in the
reference system of $(X,Y)_{2013.1}$.
So we can now compute instead the coefficients of the transformations
from these \textit{Gaia} pms-corrected positions at any given epoch
registered to $(X,Y)_{2013}$, and the observed positions of sources in
each image $(x^{\rm cor},y^{\rm cor})$.
The peculiar motions of the sources will not affect the
transformations, which will now only be affected by the positioning
random errors in the given \textit{HST} image and by errors (in both
positions and pms) in the \textit{Gaia}\,DR2 catalog.

To better expose the gain of this procedure, in Fig.\,\ref{figmat} we
show for each given image the consistency of positions measured
$(x^{\rm cor},y^{\rm cor})$ with respect to the positions of
\textit{Gaia}\,DR2. In the top panels, the \textit{Gaia}\,DR2
positions are fixed as tabulated for epoch 2015.5.  In the bottom
panels, we take advantage of \textit{Gaia}\,DR2 pms to correct the
positions of sources for their peculiar motions at the exact epoch of
the observation.
Accounting for peculiar motions of sources reduce the dispersion from
the 1-$\sigma$ of 300-50\,milli-pixels (37-6\,mas, in top panels) to a
rather uniform 15-25\,milli-pixels (1.8-3\,mas).

Once the transformations determined using sources in common between
\textit{Gaia}\,DR2 and $(x^{\rm cor},y^{\rm cor})$ are known, the
positions of all sources in all images (i.e., including those
\textit{not} present in the \textit{Gaia}\,DR2 catalog, such as
W1541$-$2250) can be transformed in the \textit{Gaia} system, and
their displacements used to measure their astrometric parameters, now
with negligible errors in the transformations.

\subsection{Stack images}
\label{stacks}
The transformations from the coordinates of each image into the
coordinates of the reference frame $(X,Y)_{2013.1}$ enable us to
create stacked images for each epoch. Stacked images offer a
representation of the astronomical scene that can be used to
independently check sources in each image.
In left panels of Fig.\,\ref{stack} we show a $\sim
20^{\prime\prime}\times6.5^{\prime\prime}$ zoom around W1541$-$2250
--from top to bottom-- for the stack from first epoch in F125W, second
epoch in F125W, and the third in F127M.
In the right panel we show the entire field of view as in first epoch,
the one with the highest signal for the BD.
Our stacked images are saved in fits format, and their headers include
(as World Coordinate System keywords) the absolute astrometric
solution of Table\,\ref{tabRDXY}.
In the electronic material provided with this work, we release these
three stacked images, one per epoch all with our astrometric solution
in their header.
Note that the $(X,Y)$ coordinates in Table\,\ref{GaiaSs} are in the
same pixel-coordinate system as these stacks.

%%%%%%%%%%%%%%%%%%%%%%%%%%%%%%%%
%%%%%%%%%%%%%%%%%%%%%%%%%%%%%%%%
%%%%%%%%%%%%%%%%%%%%%%%%%%%%%%%%

%__________________________________________________________________
\begin{figure*}
\begin{center}
\includegraphics[width=150mm]{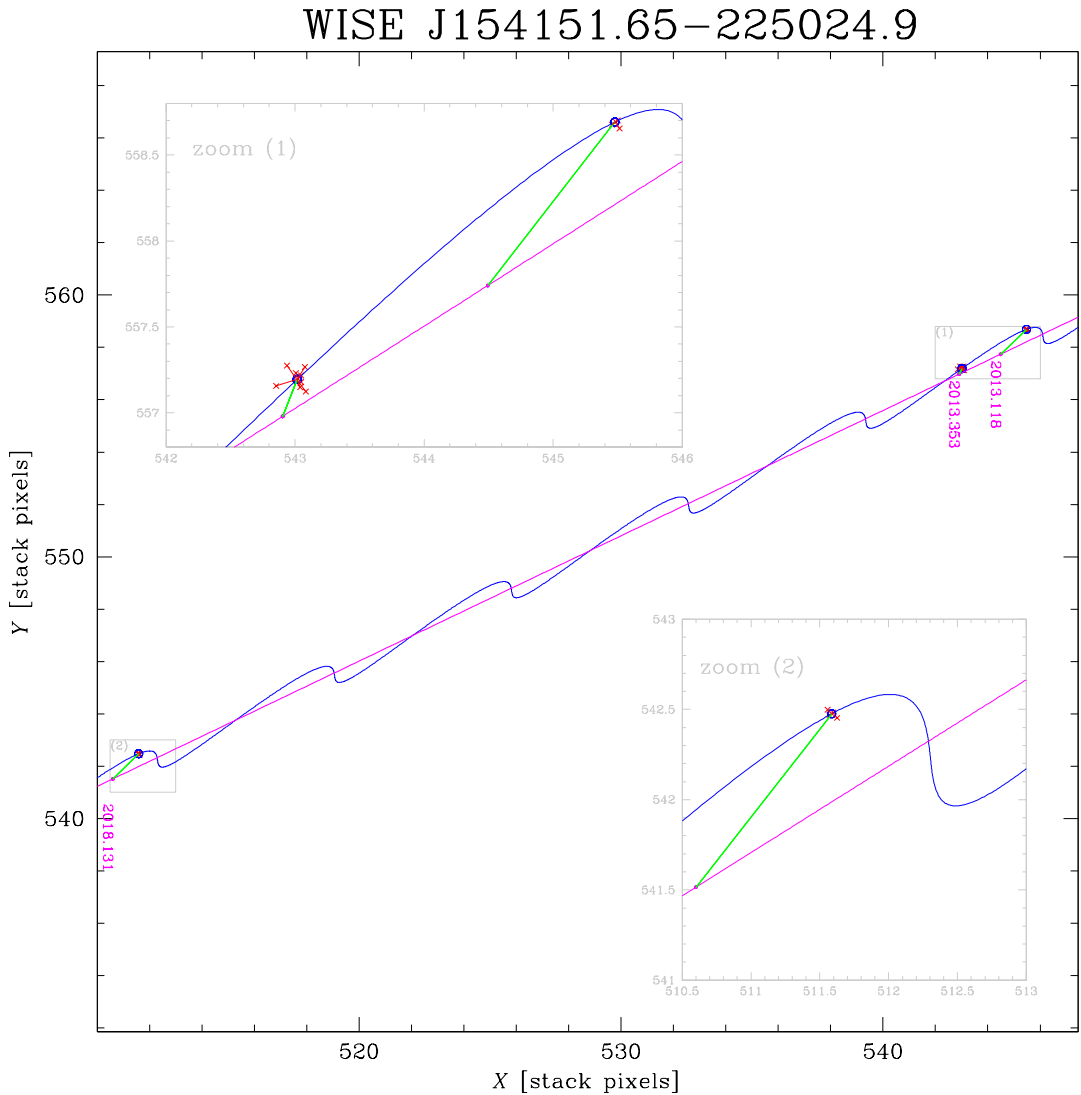}
\caption{
  Comparison of our astrometric solution (in blue) with the individual
  observed data points (red crosses) for W1541$-$2250 in the
  observational plane $(X,Y)_{2013.1}$. The three major epochs are
  indicated by labels, and two insets indicated by the gray boxes,
  with (1) and (2), show a more meaningful zoom-in of the data points.
  To better highlight the parallax component of the motion a line in
  magenta indicates an object with the same motion but at infinite
  distance. Green lines show the parallax contributions at each epoch,
  and red segment connects the individual data points with their
  expected position according to the best fit.
\label{cycloid}
}
\end{center}
\end{figure*}
%__________________________________________________________________

\section{Determination of Positions, Proper Motions and Parallax}
\label{astrpar}

From the observed 16 $\times$ 2D-data points we would like to derive
the five astrometric parameters of W1541$-$2250: its positions
$(X,Y)$, its motions $(\mu_{X},\mu_{Y})$, and most importantly the
parallax ($\pi$).
We will describe in the following the procedure followed to fit these
five parameters.

By virtue of the principle that any transformation of the
observational data degrades them, while numerical models do not, we
perform this numerical fitting process directly in the observational
plane $(X,Y)_{2013.1}$.
To predict the position of W1541$-$2250 we make use of the
sophisticated tool by U.S.\ Naval Observatory, the \textit{Naval
  Observatory Vector Astrometry Software}, hereafter NOVAS\footnote{
\texttt{http://aa.usno.navy.mil/software/novas/novas\_f/\-novasf\_intro.php}
}
(in version F3.1, Kaplan et al.\ 2011), which accounts for many subtle
effects, such as the accurate Earth orbit, perturbations of major
bodies, nutation of the Moon-Earth system, etc.
We are not interested in the absolute astrometric calculations of
NOVAS but only in the relative effects. In computing the positions we
used an auxiliary star with no motion and zero parallax (i.e., at
infinite distance), and finally compute the difference with respect to
our target.
We then use a Levenberg-Marquardt algorithm (the FORTRAN version
\texttt{lmdif} available under MINIPACK, Mor\'e et al.\, 1980) to find
the minimization of five parameters: $X,Y,\mu_{X},\mu_{Y},$ and $\pi$.

%
% X(1) =    15.697858030862172     
% X(2) =   -22.840205862620905     
% X(3) =   -902.61562101144432     
% X(4) =   -88.261065374289302     
% X(5) =    168.37633145012094     
%  
% ERRORS FROM 25000 REALIZATIONS 
% RAo_2000 =   15.69785803 +/- 5.319858747
% DCo_2000 =  -22.84020586 +/- 5.742136127
% MU_RAcsD = -902.61797440 +/- 0.3549084167
% MU_DECLI =  -88.26248465 +/- 0.3453236302
% PARALLAX =  168.38317780 +/- 2.229892366
%__________________________________________________________________
%
\begin{table}
\caption{Astrometric parameters of W1541$-$2250 in the
      ICRS. Positions are given at three relevant epochs,
      where the $^\varpi$ suffix indicates the apparent position with
      the annual parallax included.}
\center
\begin{tabular}{lcc}
%\hline
\hline
% NOT in decimal hours
%
& & \\
$\alpha_{2000.0}$ [ $^{\rm h}$ $^{\rm m}$ $^{\rm s}$ ]       &    15:41:52.28891  & $\pm$ 5.32\,mas \\
$\delta_{2000.0}$ [ $^{\circ}$ $^{\prime}$ $^{\prime\prime}$ ] & $-$22:50:24.74111  & $\pm$ 5.74\,mas \\
& & \\
$\alpha_{2000.0}$ [degrees] &   235.46787051 & $\pm$ 5.32\,mas \\
$\delta_{2000.0}$ [degrees] & $-$22.84020586 & $\pm$ 5.74\,mas \\
& & \\
$\alpha_{2013.1}^\varpi$ [degrees] &   235.4643508 & $\pm$  1.0  \,mas \\
$\delta_{2013.1}^\varpi$ [degrees] & $-$22.84053754 & $\pm$ 1.3  \,mas \\
& & \\
$\alpha_{2015.5}$ [degrees] &   235.46365494 & $\pm$ 2.3  \,mas \\ % ok
$\delta_{2015.5}$ [degrees] & $-$22.84058569 & $\pm$ 1.2  \,mas \\ % ok
& & \\
$\alpha_{2015.5}^\varpi$ [degrees] &   235.46362187 & $\pm$  3.8  \,mas \\ % ok
$\delta_{2015.5}^\varpi$ [degrees] & $-$22.84058063 & $\pm$  1.4  \,mas \\ % ok
& & \\
$\mu_{\alpha\cos{\delta}}$ [mas yr$^{-1}$]   & $-$902.62 & $\pm$ 0.35 \\
$\mu_{\delta}$ [mas yr$^{-1}$]  &  $-$88.26   & $\pm$ 0.35 \\
& & \\
%
% error of 0.01324 i.e. ~1.32%
$\pi$ [mas]  & 168.38     & $\pm$ 2.23 \\
$\varpi=\pi+0.2$  [mas]  & 168.58     & $\pm$ 2.23 $\pm$0.4 \\
\hline
\end{tabular}
\label{tabASTR}
\end{table} 
%__________________________________________________________________
%

%%%
%%%

%
Our final astrometric solution is given in Table\,\ref{tabASTR} and
shown in Fig.\,\ref{cycloid}.
To assess the uncertainties of our solution we perform 25\,000
simulations where, to the expected positions $(X,Y)$ from our best-fit
astrometric solution, we added random errors following Gaussian
distributions with dispersion derived from the observed data of
W1541$-$2250 for each of the four filter/epoch combinations (i.e.,
F125W@2013.1, F105W@2013.3, F125W@2013.3, and F127M@2018.1).

Our astrometric parameters agree well with the two best estimates in
literature, those by Tinney et al.\ (2014) and Beichman et
al.\ (2014), and represent a significant improvement. Furthermore, our
solution surely does not suffer of any of the usual ground-based
atmospheric effects and relies on the homogeneous data, therefore
providing an important confirmation.\\

\begin{figure}
\begin{center}
\includegraphics[width=88mm]{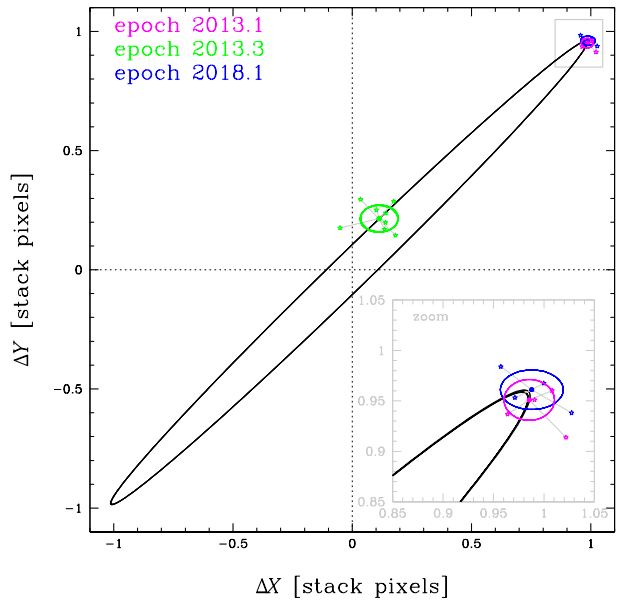}
\caption{
  Our solution for the parallax ellipse in the $(X,Y)_{2013.1}$
  coordinate system.  Individual \textit{HST} data-points are
  indicated with star-symbols, which are connected with small segment
  to their expected position according to our best fit.  Smaller
  ellipses in magenta, green, and blue, indicate the 1-$\sigma_{X,Y}$
  of individual data points within the first, second and third epoch,
  respectively.  Note how ellipses are significantly smaller for the
  first and third epoch (0.025-0.020 and 0.031-0.020\,WFC3/IR-pixels,
  i.e., 3.0-2.4 and 3.8-2.4\,mas), compared to the second epoch
  (0.079-0.056\,pixels, i.e., 9.5-6.8\,mas).  However, second epoch
  has twice as many images than each of the other two, therefore, as
  the error on the average $\overline{\sigma}$ scales as
  $\sim\sigma/\sqrt{n-1}$ (with $n$ equal to the number of images),
  the net worsening for this epoch is at most a factor 2. An inset in
  gray, zoom-in at a busy location around the maximum parallax
  elongation, marked by a gray box.
\label{Ell}
}
\end{center}
\end{figure}

In addition to Fig.\,\ref{cycloid} and its insets, in Fig.\,\ref{Ell}
we show also the parallax ellipse along with \textit{HST} measurements
\textit{[proper motion subtracted].}  This representation better
reveals the sampling of the parallactic motion which, with only three
epochs, could be problematic.

We note few things in this figure. First, the parallax ellipse is
extremely flattened, as an obvious consequence of the extremely
low-latitude ecliptic coordinates for W1541$-$2250, i.e.,
$(\ell,b)=(\sim238^\circ,\sim-3^\circ)$.

Second, we know that the first (2013.1) and the last epoch (2018.1)
were collected almost exactly 5\,yrs apart; meaning that they were
collected almost exactly at the same phase of the year.

Third, we know also these two epochs to be taken at one of the maxima
of the parallax elongation, and indeed, their positions on the
parallax ellipse agree with eachother, and with the position of the
maximum.  These two epochs also have the greatest accuracies (thanks
to the filter/exposure-time combination) and have the largest possible
time base-line among the epochs.
This is a great advantage to derive the astrometric parameters, as
these two epochs alone could essentially fix the proper motions of
W1541$-$2250 with great accuracy.\footnote{
It is interesting how a crude calculation based on simply summing in
quadrature the uncertainties on these two epochs divided by the time
base-line
($\sim$3\,mas$/\sqrt{4-1}\times\sqrt{2}/5$\,yr\,$\simeq$\,0.49\,mas)
provide a proper motion uncertainty consistent with our best fit in
Table\,\ref{tabASTR} (0.35\,mas), which was derived using also epoch
2013.3.
} 
The determination of the parallax therefore, essentially relies in the
second epoch (2013.3), which thankfully was collected at a
significantly different phase of the year allowing us to well
constrain this astrometric parameter.\footnote{
We note that Beichman et al.\ (2014) had at disposal only two out of
the three \textit{HST} epochs analayzed here.
}
Unfortunately the second epoch has also the lowest accuracy of the
three and therefore ultimately sets the limits in the accuracy of the
here derived parallax for W1541$-$2250.

%__________________________________________________________________
%__________________________________________________________________

%__________________________________________________________________
%
\begin{table*}
  \caption{
    List of works in the Literature providing astrometric parameters for W1541$-$2250.}
  \center
\begin{tabular}{lccccr}
\hline
work  & $\mu_{\alpha^*}\pm\sigma\mu_{\alpha^*}$ & $\mu_\delta\pm\sigma\mu_\delta$ & $\pi\pm\sigma_\pi$ & $d$ & source\\
\#. authors (date) & [mas\,yr$^{-1}$] & [mas\,yr$^{-1}$] & [mas] & [pc] & facilities \\
\hline
   1. Cushing et al.\ (2011)    & ...          & ...           & ...           & 8.1 (8.1-8.9) & Magellan+\textit{WISE} \\ % spectroscopic distance
   2. Kirkpatrick et al.\ (2011)& ...          & ...           & $351\pm108$   & 2.2-4.1       & Magellan \\
   3. Kirkpatrick et al.\ (2012)& ...          & ...           & $238$         & 4.2$^{*}$      & \textit{WISE} \\ % trig parallax (not used), spectrophotometric ("adopted") distance 
   4. Dupuy \& Kraus (2013)     & $-870\pm130$ & $-13\pm58$    & $74\pm31$     & $14^{+10}_{-4}$  & \textit{Spitzer}\\ % MCMC distance gives $14^{+10}_{-4}$
   5. Marsh et al.\ (2013)      & $-983\pm111$ & $-276\pm116$  & $-21\pm94$    & $> 6.0$       & \textit{WISE}+\textit{Spitzer}+Magellan+CTIO \\
   6. Tinney et al.\ (2014)     & $-894.7\pm4.2$ & $-87.7\pm4.7$ & $175.1\pm4.4$ & ...         & Magellan \\
   7. Beichman et al.\ (2014)   & $-857\pm12$  & $-87\pm13$    & $176\pm9$     & $5.7\pm0.3$   & Keck+\textit{Spitzer}+\textit{HST} \\
   8. Martin et al.\ (2018)$^\dagger$
                                & $-895.05\pm4.68$ & $-94.73\pm4.66$ & $167.05\pm4.19$ & $5.99^{+0.154}_{-0.147}$ & \textit{Spitzer} \\
   9. \textbf{this work}        & $-902.62\pm0.35$ & $-88.26\pm0.35$ & $169\pm2$ & $5.9\pm0.1$ & \textit{HST+Gaia} \\
\hline
 \multicolumn{6}{l}{$^*$ spectro-photometric distance, to which they refer as 'adopted distance'}\\
 \multicolumn{6}{l}{$^\dagger$ while our article was still under review}\\
\end{tabular}
\label{lit}
\end{table*} 
%__________________________________________________________________
%

%%%%%%%%%%%%%%%%%%%%%%%%%%%%%%%%%%%%%%%
%
\subsection{The Absolute Parallax}
%
%%%%%%%%%%%%%%%%%%%%%%%%%%%%%%%%%%%%%%%

The transformations of coordinates described in previous sections were
computed with respect to \textit{Gaia}\,DR2 sources, which are not at
infinite distance.
Therefore the relative parallax we have derived in
Sect.\,\ref{astrpar} (which we indicate with $\pi$) of W1541$-$2250 is
with respect to the most distant objects in the field and therefore
$\pi$ is only a lower limit on the absolute parallax (which we
indicate with $\varpi$).
\textit{Gaia}\,DR2 catalog potentially gives the possibility to
correct the positions of sources --at any give epoch-- not only for
individual pms, but also for the amplitude of the parallax at any
given phase of the year.
However, given the accuracies in this work, it will be sufficient to
correct for the clipped mean parallax of the sources used to compute
the parameter of the transformation given in Table\,\ref{tabASTR}.
Twelve of the 15 stars used to compute the transformation have a
parallax consistent with zero, and only two have a significant
parallax of about $\sim$1\,mas. Only one of the reference sources,
\texttt{Gaia\,DR2\,\#6240155613540444800}, had a significant large
parallax of 3.857$\pm$0.606\,mas, and was rejected.
The computed average parallax of the remaining 14 sources is
0.3$\pm$0.2\,mas, while their median is 0.16\,mas with a
semi-interquartile of 0.35\,mas.
Given the uncertainties, we adopt 0.2$\pm$0.4\,mas as correction from
relative to absolute parallax.
In Table\,\ref{tabASTR} we also give the value derived for $\varpi$
employing this correction. The $\varpi$ and uncertainties translate in
an estimated distance for W1541$-$2250 of $d=5.9\pm0.1$\,pc.
The uncertainty in the estimated relative parallax of W1541$-$2250
amounts to $\sim$2.3\,mas, making the correction from relative to
absolute an unnecessary calculation in this case.

Future applications of this method to cases that would actually need
higher accuracy will require to calculate ---for each of the reference
sources--- the exact positions corrected for the parallax (in addition
to corrections for pms), before computing the coordinate
transformations.
%

%%%%%%%%%%%%%%%%%%%%%%%%%%%%%%%%%%%%%%%
%
\section{Conclusions}
%
%%%%%%%%%%%%%%%%%%%%%%%%%%%%%%%%%%%%%%%

In this paper we employed three archival and proprietary \hst\, epochs
to derive an independent and robust estimate of the astrometric
parameters of WISE\,J154151.65$-$225024.9, one of the only 23
confirmed Y BDs (as September 14$^{\rm th}$ 2018), and the fifth
closest.\footnote{
\texttt{https://sites.google.com/view/ydwarfcompendium/home}
}

We have developed a procedure that makes use of the \textit{Gaia}\,DR2
catalog of positions, proper motions and parallaxes, to minimize the
errors in the transformations that used to limit traditional relative
imaging-astrometry with narrow-field cameras.
We gave here an extensive explanation and many details of our
procedure, with the intent of using it in future articles that will
refer to the present work. Indeed, future papers will be focused on
the science (mainly based on the derived multi-band photometry) and
will minimize description of the technicalities to derive the
distances.

The derived astrometric parameters for W1541$-$2250 are summarized in
Table\,\ref{tabASTR} and represent sizable improvements over recent
determinations in the literature which are summarized in
Table\,\ref{lit}.
Worth to notice that the here derived relative parallax of $\pi =
168.38 \pm 2.23$\,mas (i.e., with an accuracy better than 1.3\%) is
well in agreement (at the 1.5\,$\sigma$-level) with the best available
(and completely independent) estimate by Tinney et al.\ (2014) ($\pi =
175.1 \pm 4.4 $\,mas), which was based on ground-based observations.
Ground-based observations are generally afflicted by atmospheric
effects and gravity-induced flexes in the telescope/camera structures
that generate systematic errors often difficult to correct,
particularly when assessing the astrometry of faint and red sources
(much redder than reference field stars).
Therefore, the homogeneous space-based \hst\, data provide an
important update and confirmation of those previous estimates for the
W1541$-$2250 parallax.

Applying a \textit{Gaia}\,DR2 correction from relative ($\pi$) to
absolute ($\varpi$) parallax we found $\varpi$=169$\pm$2\,mas,
corresponding to a distance of 5.9$\pm$0.1\,pc for W1541$-$2250,
marginally larger than previously estimated,
but in agreement with the most recent determination by Martin et
al.\ (2018, accepted paper in press, private communication) based on
\textit{Spitzer} data.
\section*{Acknowledgments}
We are grateful  to Chas Beichman for his prompt  and competent review
of  our work.  His careful  reading  and comments  had contributed  to
improve our manuscript.
We thank Dr.\ Emily Martin and Dr.\ Davy Kirkpatrick for sharing
results from their accepted --but--still--not--public-- article.
We take the chance also to sincerely thank the editors and MNRAS
personnel, for their constant: kindness, helpfulness, efficiency,
competence and good work.
This work is based on observations with the NASA/ESA Hubble Space
Telescope, obtained at the Space Telescope Science Institute, which is
operated by AURA, Inc., under NASA contract NAS 5-26555.
This work makes also use of results from the European Space Agency
(ESA) space mission Gaia. Gaia data are being processed by the Gaia
Data Processing and Analysis Consortium (DPAC). Funding for the DPAC
is provided by national institutions, in particular the institutions
participating in the Gaia MultiLateral Agreement (MLA). The Gaia
mission website is \texttt{https://www.cosmos.esa.int/gaia}. The Gaia
archive website is \texttt{https://archives.esac.esa.int/gaia}.
This research has benefitted from the Y Dwarf Compendium maintained by
Michael Cushing at
{\texttt{https://sites.google.com/view/ydwarfcompendium/}.

\noindent

%
%%
%%%
%%%%%%%%%%%%%%%%%%%%%%%%%%%%%

%%%%%%%%%%%%%%%%%%%%%%%

\label{lastpage}

%%%%%%%%%%%%%%%%%%

\end{document}